\documentclass[aip,preprint]{revtex4-1}

\usepackage{graphicx}

\draft

\begin{document}

\title{Atomic-scale effects behind structural instabilities in Si
  lamellae during ion beam thinning}

\author{E. Holmstr\"om$^{\scriptsize{\textrm{1,2,}}}$\footnote{Corresponding
    author, email: eero.holmstrom@helsinki.fi},
  J. Kotakoski$^{\scriptsize{\textrm{1,}}}$\footnote{Current address:
    Department of Physics, University of Vienna, Boltzmanngasse 5,
    1090 Wien, Austria},
  L. Lechner$^{\scriptsize{\textrm{3,}}}$\footnote{Current address:
    Carl Zeiss NTS GmbH, Carl-Zeiss-Stra\ss e 56, 73447 Oberkochen,
    Germany}, U. Kaiser$^{\scriptsize{\textrm{3}}}$, and
  K. Nordlund$^{\scriptsize{\textrm{1,2}}}$\\
  $^{\scriptsize{\textrm{1}}}$\emph{Department of Physics, University
    of Helsinki, P.O. Box 64, Helsinki FIN 00014, Finland}\\
  $^{\scriptsize{\textrm{2}}}$\emph{Helsinki Institute of Physics,
    P.O. Box 64, Helsinki FIN 00014, Finland}
  \\$^{\scriptsize{\textrm{3}}}$\emph{Universit\"at Ulm, Central
    Facility of Electron Microscopy, Albert Einstein Allee 11 89069
    Ulm, Germany}}

\date{\today}

\begin{abstract}

  The rise of nanotechnology has created an ever-increasing need to
  probe structures on the atomic scale, to which transmission electron
  microscopy has largely been the answer. Currently, the only way to
  efficiently thin arbitrary bulk samples into thin lamellae in
  preparation for this technique is to use a focused ion beam
  (FIB). Unfortunately, the established FIB thinning method is limited
  to producing samples of thickness above $\sim$20~nm. Using atomistic
  simulations alongside experiments, we show that this is due to
  effects from finite ion beam sharpness at low milling energies
  combined with atomic-scale effects at high energies which lead to
  shrinkage of the lamella. Specifically, we show that attaining
  thickness below 26~nm using a milling energy of 30~keV is
  fundamentally prevented by atomistic effects at the top edge of the
  lamella. Our results also explain the success of a recently proposed
  alternative FIB thinning method, which is free of the limitations of
  the conventional approach due to the absence of these physical
  processes.

\end{abstract}


\maketitle

%\section{INTRODUCTION}

Since atomic-scale building blocks determine functionality in
nanotechnology, understanding physical processes at this scale is
crucial for the discipline. One of the most important means for
achieving this knowledge is aberration-corrected high-resolution
transmission electron microscopy (AC-HRTEM),\cite{Rose1990,Haider1998}
a method which allows determining the positions of atomic columns of a
structure with accuracy higher than interatomic distance in
solids.\cite{Jia2003,Zhang2009} HRTEM requires thin, unbent samples
which ideally should contain no amorphous surface layer. Moreover, in
nanotechnology, it is mandatory to achieve sample preparation with a
sub-$\mu$m precision in all dimensions. These standards cannot be
reached using traditional mechanical preparation methods such as
mechanical milling, and thus the focused ion beam (FIB) technique,
where the sample is milled into a thin lamella through sputtering,
needs to be used instead.

The current development in AC-HRTEM is to lower the voltage from the
usual 200--300~kV to well below
100~kV,\cite{Meyer2011,Chuvilin2009,Suenaga2009,Huang2011,Sasaki2011,Kaiser2011}
in order to minimize displacement damage in fragile
samples.\footnote{For example, voltages below 50~kV are required to
  prevent displacement of atoms at graphene edges or next to
  vacancies.\cite{Kotakoski2012}} So far, predominantly
low-dimensional materials such as
graphene\cite{Meyer2011,Huang2011,Kaiser2011}, hexagonal boron
nitride\cite{Chuvilin2009}, and carbon nanotubes\cite{Suenaga2009}
have been studied at low electron beam voltages by HRTEM. In the
future, however, there will be a demand to extend these low-voltage
studies to conventional materials, which similarly suffer from
knock-on damage at higher voltages. This trend of lowering voltage
entails an increasingly stringent requirement for sample thickness,
which must not surpass the extinction length of the electron beam
($\sim$few nm). However, the conventional FIB thinning method is
limited to thicknesses above $\sim$20~nm. The dominant effect behind
this limitation is the shrinkage of the lamella in height during
milling: Thinning a lamella down to a thickness of 20~nm may induce
several $\mu$m of shrinkage in the vertical
direction.\cite{Kang2010} The underlying physical reasons for this
observation are currently not known, because an atomic-scale
description of the thinning process has hitherto not been
presented. Additionally, the conventional method is hampered by
specimen warping, an uneven thickness profile, and often heavy
amorphization.

Although irradiation effects in Si have been studied for decades, the
work has concentrated on bulk Si
samples.\cite{Cha97,Pos00b,Nor97f,Ric07} Moreover, earlier
computational work on FIB processing of similar nanosystems has been
restricted to simulation setups which severely reduce the
applicability of the obtained results to the problem addressed in the
present study. On the one hand, some studies have employed less
detailed models based on Monte Carlo simulation of sputtering and
redeposition on a three-dimensional grid.\cite{Katardjiev1994,Kim2007}
Unfortunately, such models lack real atomic resolution and hence do
not account for, {\it e.g.}, amorphization, which alone renders them
powerless for the topic at hand. On the other hand, previous molecular
dynamics (MD) simulations motivated by FIB milling have employed
geometries restricted to a single surface with no deformation of the
sample allowed,\cite{Russo2008,Pastewka2009,Giannuzzi2009} thus
yielding comparatively limited insight into the process.  To uncover
the mechanisms responsible for the deleterious effects on the lamella
under conventional FIB thinning, we performed a classical MD study of
the FIB thinning process in conjunction with experiments. Using the
findings from our experiments and simulations, we show that the
empirically observed limit of 20~nm for the conventional thinning
method is a result of geometric sputtering effects due to the finite
sharpness of the ion beam, dominant at low energies ($\sim$1~keV), and
of atomic-scale effects at the very top part of the lamella,
significant at high energies ($\sim$30~keV). We further explain why
these mechanisms are not present in a recently proposed {\it
double-tilt} method,\cite{Lechner2011} which thus allows the thinning
process to proceed beyond the conventional limit.

%\section{RESULTS AND DISCUSSION}

\begin{figure} [ht]
\includegraphics[width=0.5\textwidth]{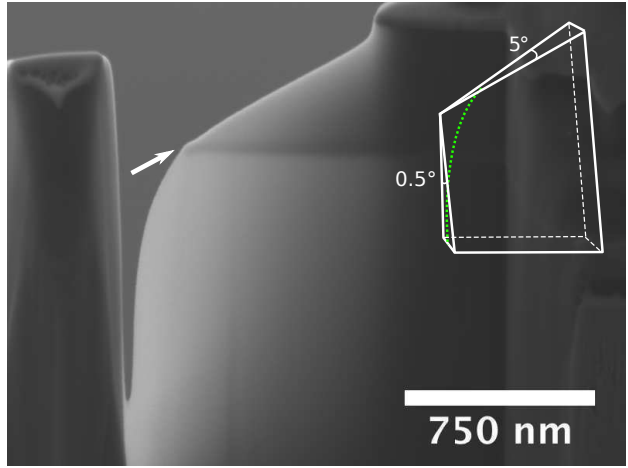}
\caption{SEM image of the double-wedge lamella. The horizontal angle is
  5$^\circ$, the vertical angle is 0.5$^\circ$, making the lamella
  thinner from bottom to top and from right to left. Thin areas appear
  lighter in secondary electron imaging mode. The thin top edge is
  indicated by the white arrow. The dropping edge indicates that there
  is a thickness-dependent change in sputtering rate; otherwise a
  straight edge would be expected, as illustrated in the schematic in
  the upper right-hand corner of the figure (the green dotted line
  indicates the dropping edge, the white lines indicate the expected
  shape).
  \label{fig:lamella_wedge}}
\end{figure}

The experimental setup and the conventional FIB thinning method are
described in Appendix A. The effect of ion beam tails, discussed
below, is demonstrated in Appendix B. Here, we focus first on studying
experimentally the deleterious shrinkage effect on a lamella during
FIB thinning. In principle, creating a wedge-shaped lamella would show
shrinkage versus thickness in a straightforward experiment. However,
the structural stability of such a geometry is reduced fast during
thinning, and warping of the sample occurs. Therefore, we instead
thinned a co-planar Si lamella in a double-wedge configuration
(Fig.~\ref{fig:lamella_wedge}): Material was removed from one side by
milling at a horizontal tilt of 4$^\circ$ and from the other side at
1$^\circ$ but with a vertical rotation of 0.5$^\circ$. The resulting
lamella thus gets gradually thinner from bottom to top and from right
to left with the very thin areas restricted to the top edge to limit
warping. Milling parameters for the fabrication were identical to the
single-edge experiment of Appendix B. For a given material and set
beam conditions the absolute thickness of the lamella can be estimated
from scanning electron microscope (SEM) image
contrast.\cite{Salzer:2009jq} In the secondary electron SEM image of
Fig. \ref{fig:lamella_wedge}, the thickness gradient of the lamella
manifests in a contrast gradient with the thin areas appearing
lighter.

Based on purely the milling geometry, the thin top-edge of the created
double-wedge lamella should be straight, as shown in the schematic
illustration of Fig.~\ref{fig:lamella_wedge}. Instead, we observe a
strongly falling edge indicating that the removal of lamella material
in the vertical direction is greatly increased when the thickness
drops below a certain threshold. Furthermore, very thin areas ($<
10$~nm) are notably absent in the wedge lamella. In fact, we could not
find any experimental conditions that would have allowed us to create
a very thin Si edge using 30 keV Ga ions. We tried shortening the line
dwell time/milling depth to reduce the corner rounding -- to no
avail. Reducing the beam current and thus sharpening the beam did not
prove a viable option either since the resulting milling times were
excessive, leading to problems from stage drift. Consequentially, we
can unfortunately not draw any quantifiable conclusions from the
thinning experiments except that we were not able to create very thin
areas, and that there is a thickness-dependent vertical shrinkage
effect for the thinnest parts of the structure. \footnote{Similarly,
we find that it seems not possible to create tips, {\it e.g.}, from
tungsten, with a radius smaller $\sim$ 20 nm.}

\begin{figure} [ht]
\includegraphics[width=0.3\textwidth]{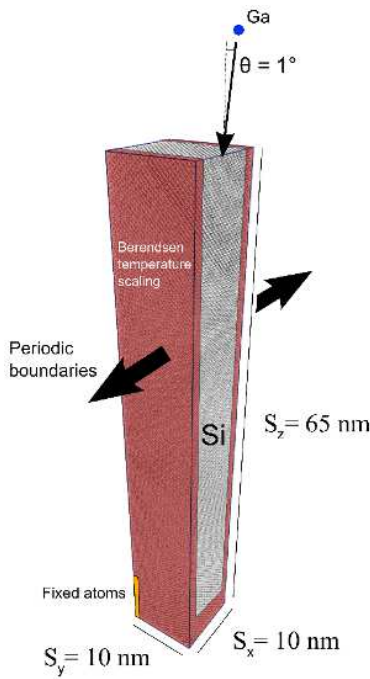}
\caption{The simulation setup for a 10~nm thick lamella top edge, as
  visualized with {\scshape ovito}.\cite{OVITO} The total number of
  atoms in this structure is 310 000.\label{fig:setup}}
\end{figure}

In order to understand the unexpected shrinkage of the lamella below the
threshold thickness, an atomic-level description of the irradiation
process is required. Our MD simulation setup, designed for studying
ion irradiation effects on the top edge of a short section of a thin
Si lamella, is visualized in Fig. \ref{fig:setup}. In all simulations,
the studied system size was 65~nm in the vertical ($z$) direction and
10~nm in width ($x$-direction), periodic boundaries being applied in
the latter dimension to mimic a much wider lamella edge. The beam,
assuming a Gaussian distribution with a standard deviation of 2.5~nm
in both the $x$ and $y$-direction, was directed at the edge of the
lamella at an angle of $1^\circ$. The thickness of the edge was
controlled by the size of the system in the $y$-direction ($S_y \in
[3, 10]$~nm). The simulation setup is further detailed in Appendix C,
and the performed runs are described in Appendix D.

\begin{figure} [ht]
\includegraphics[width=0.6\textwidth]{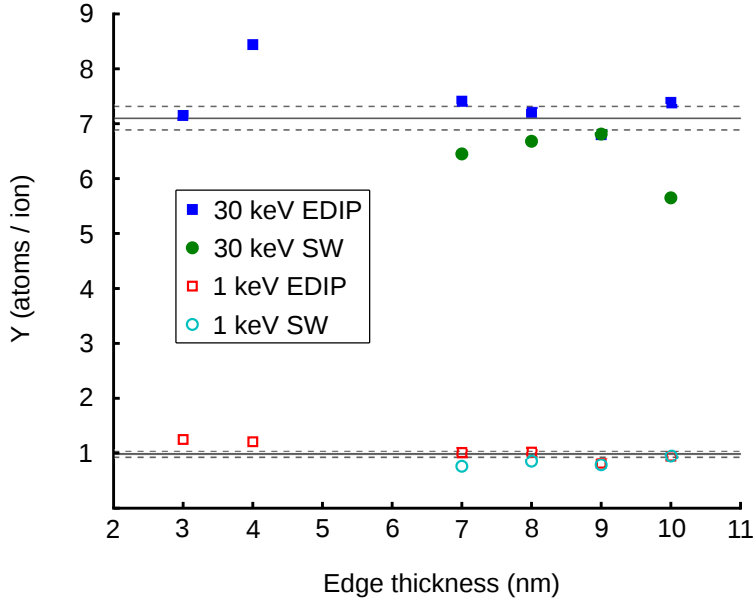}
\caption{Sputtering yield $Y$ as a function of edge thickness up to a
  dose of 500 ions. The averages and the respective errors are shown
  by the horizontal solid and dotted lines, respectively. Estimated
  uncertainties for $Y$, obtained from the fitting, are contained
  within the markers. \label{fig:sputtering_yield}}
\end{figure}

In Fig.~\ref{fig:sputtering_yield}, we plot sputtering yield $Y$ as
a function of edge thickness for 1~keV and 30~keV Ga$^+$ irradiation
as given by two different Si--Si interaction potentials (Appendix
C). The data is clearly grouped by the beam energy: The thickness of
the system seems to have no systematic effect on the results. On
average, the 30~keV beam produces a $Y$ of $7.1 \pm 0.2$ atoms/ion,
whereas the 1~keV beam gives $0.98 \pm 0.05$ atoms/ion, as averaged
over all system sizes and potentials. To better understand whether the
irradiation energy leads to other differences beyond those in $Y$
during the milling, we carried out an extensive analysis of different
physical effects occuring during the process. For all simulated cases,
the initial positions of sputtered atoms were determined after each
irradiation event. Secondly, the center of mass of the system in the
$y$-direction $CM_y$ was calculated as a function of the
$z$-coordinate. Thirdly, the degree of amorphization of each system
was determined likewise as a function of $z$ using structure factor
analysis.\cite{Nor96d}

\begin{figure} [ht]
  \includegraphics[width=0.9\textwidth]{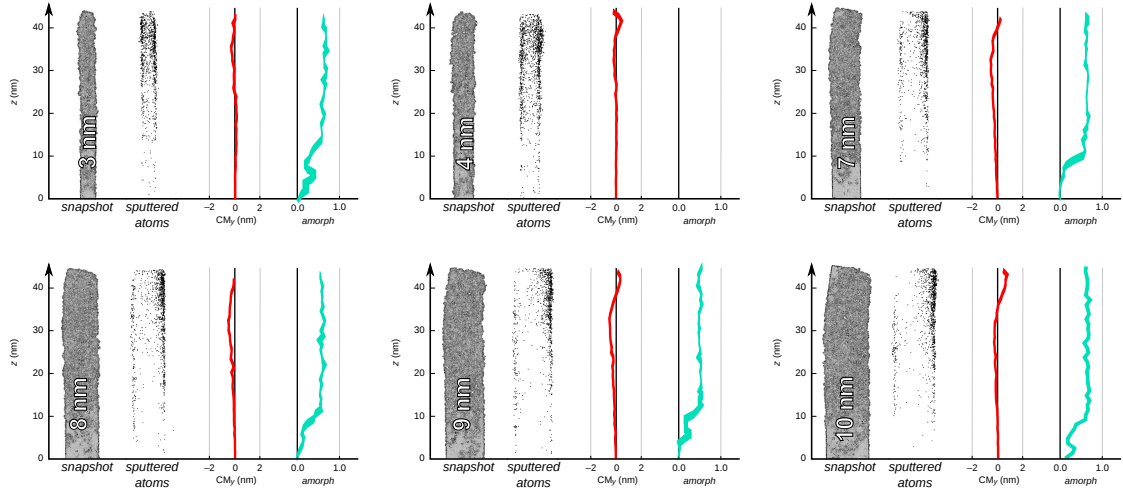}
  \caption{Results for the edges of all studied thicknesses after a
    dose of 500 ions at 30~keV. For each case, we show the snapshot of
    the final structure along the $x$-direction, the positions of
    sputtered ions, the center of mass $CM_y$, and the degree of
    amorphization. Only the upper part of the structure is shown, as
    this is where all interesting effects
    appear.\label{fig:end_states_EDIP}}
\end{figure}

Our analysis shows that the 1~keV beam results in the smoothening of
the initially abrupt upper corner of the edge, which would lead to
milling of the edge upon continued irradiation. The corner is
amorphized, and sputtering is mainly from the impact region. However,
the effect of the 30~keV beam, as presented in
Fig.~\ref{fig:end_states_EDIP}, is more interesting. At this beam
energy, the $\ge$7~nm structures bend toward the edge onto which the
beam is aimed. This is verified by the diagrams showing the center of
mass in the $y$-direction as a function of the edge height, which
clearly reflect what is seen in the snapshot of the final state of
each system. Also, the apparent amorphization seen in the snapshots is
confirmed by the quantitative amorphization analysis. Sputtering is
induced over the entire front face of the system and somewhat at the
back face also. For edges $<$7~nm, however, almost no bending is
observed at the dose of 500 ions. Instead, the system simply shrinks
vertically as a result of the ion bombardment. Correspondingly,
sputtering for the thinnest structures is no longer predominantly off
the front face but also significant off the back of the structure.

\begin{figure} [ht]
\includegraphics[width=0.6\textwidth]{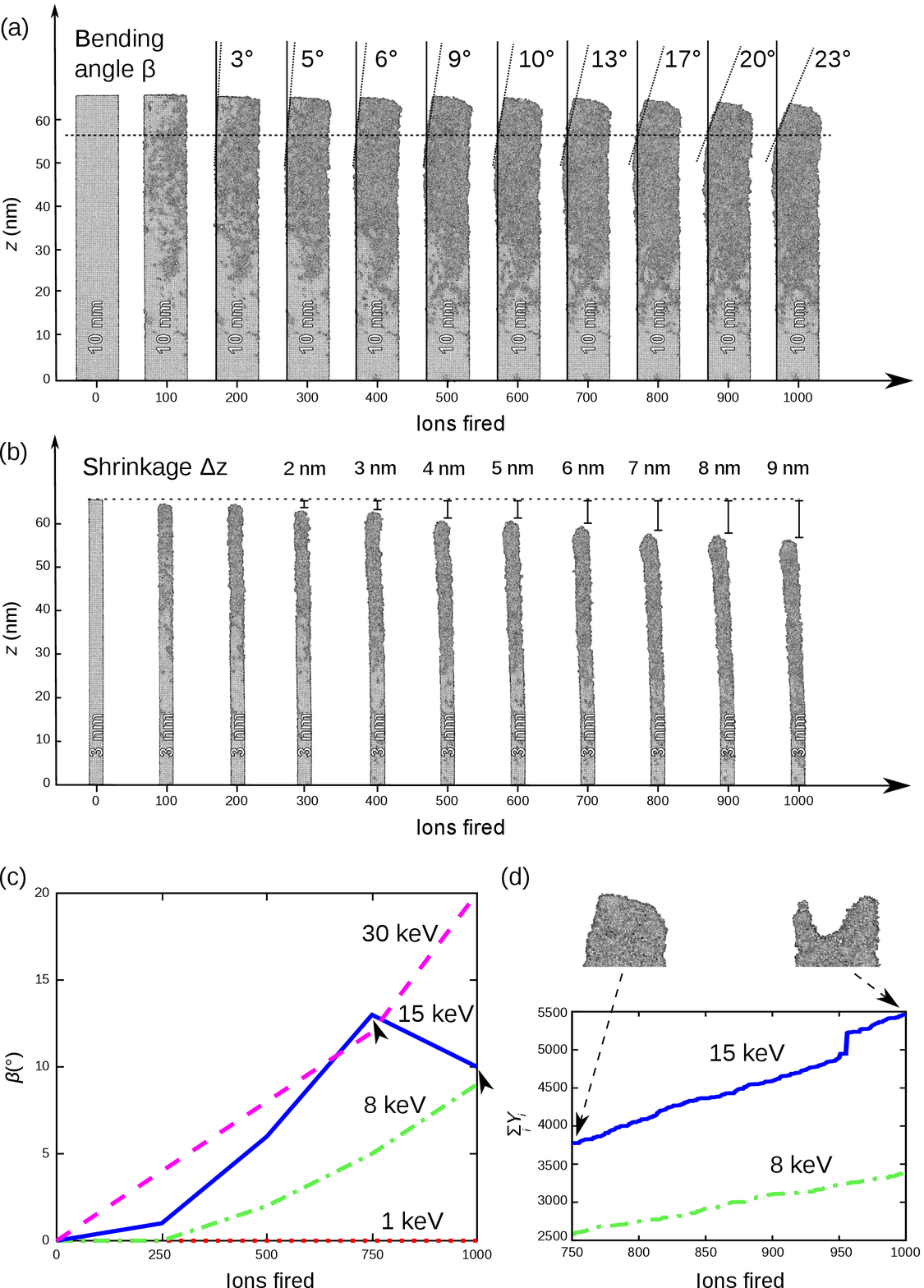}
\caption{Evolution of the (a) 10 and (b) 3~nm edges as a function of the dose of a 30~keV beam.
The 10~nm thick system exhibits mainly bending, whereas the 3~nm one exhibits predominantly shrinking. The
  dashed horizontal line in (a) depicts the hinge around
  which the bending occurs. The snapshot of the structure is a sideview
  along the $x$-direction. (c) The bending angle of the
  10~nm edge at beam energies between 1~keV and 30~keV as a function of
  dose. (d) Cumulative sputtering yield $\sum_i Y_i$ for the 10~nm edge as a function of
  the dose between 750 to 1000 ions at 8 and 15~keV. The sharp
  increase close to 950 ions for the 15~keV case explains the sudden
  drop in the bending angle seen in panel (c) (as indicated by the
  arrows): An exceptionally dense region of deposited energy by one of
  the incoming ions craterizes the top of the structure and thus
  destroys the forward-bent shape (see the snapshots), leading to the
  experimentally observed shrinkage via the bending mechanism.
  \label{fig:struct_vs_dose}}
\end{figure}

Based on the results presented above, it would seem that the 30~keV
beam induces mainly bending and modest shrinking in the thicker edges
($\ge$7~nm) and predominantly shrinking in the thinnest edges. This
inference is supported by studying the evolution of the atomic
structure of the 10 and 3~nm edges as a function of the 30~keV
irradiation dose, as presented in Figs \ref{fig:struct_vs_dose}(a,b),
respectively. As the dose increases, the 10~nm system bends toward the
corner where the beam is incident, whereas the 3~nm system shrinks by
nearly 10~nm in height as the dose is brought to 1000 ions. The
explanation for these two distinct modes of behavior can be found by
considering the positions of sputtered atoms in each case and taking into
account the resulting surface tension. As noted above and as seen in
Fig. \ref{fig:end_states_EDIP}, sputtering for the $\geq$7~nm edges
occurs mainly off the front face of the structure. Therefore, the
system will relax, {\it i.e.}, minimize its free energy to a local
minimum, by contracting the front face in the vertical dimension,
hence pulling the structure into a forward-bent
position. Correspondingly, when atoms are sputtered evenly off both
the front and back faces, as for the $<$7~nm edges, the bending
behavior is taken over by the shrinkage of the system in the vertical
direction, as the structure relaxes by contracting along the entire
thickness.

Note that the above-described bending is not the same phenomenon as
the deleterious warping observed experimentally in the sample during
conventional FIB thinning. The empirical warping happens in the
opposite direction relative to the beam, and throughout the bulk of
the lamella. Perhaps counterintuitively, the bending as described by
our simulations will actually appear as shrinkage of the lamella top
in the experiments. This is because the tilted structure is etched
away more efficiently by the incoming ions than the originally
straight-standing one. Obviously, this erosion process is very
difficult to observe experimentally, and has therefore not been
detected before. One example of how the erosion might proceed is
presented below in Fig.~\ref{fig:struct_vs_dose}(d). Here a large
density of deposited energy on the top of the bent structure causes
hundreds of atoms to sputter with a single ion hit and leads to the
subsequent straightening of the lamella edge. For the extreme case of
lamella edges $<$7~nm, sputtering from both the front and back faces
simply constitutes another mechanism of shrinkage.

Focusing on the described bending of the edge in more detail, it can
be seen in Fig. \ref{fig:struct_vs_dose}(c) that an 8~keV ion beam is
energetic enough to induce bending in the 10~nm structure, whereas at
1~keV the edge is left standing upright. Looking at the bent
structures in Figs \ref{fig:end_states_EDIP} and
\ref{fig:struct_vs_dose}(a), it can be discerned that the top part of
the bent structure is as if bent around a hinge at a well-defined
height on the back face of the structure, as illustrated explicitly by
the horizontal line in Fig. \ref{fig:struct_vs_dose}(a). This point is
roughly around the area where the amorphization starting from the
front face of the structure reaches the back face, which is where the
originally rigid crystalline back wall gives way to the contracting
force on the front face of the system. Such a mechanism is functional
only when the system is thin enough to allow for the amorphization to
proceed all the way through the structure. For sufficiently thick
structures this does not happen. We emphasize that this does not
require a complete amorphization of the structure. Instead, it is
enough if the damage reaches the back face at one location. This may
happen outside the collision cascade, as was shown
recently.\cite{Pothier2011}

\begin{figure} [ht]
  \includegraphics[width=0.7\textwidth]{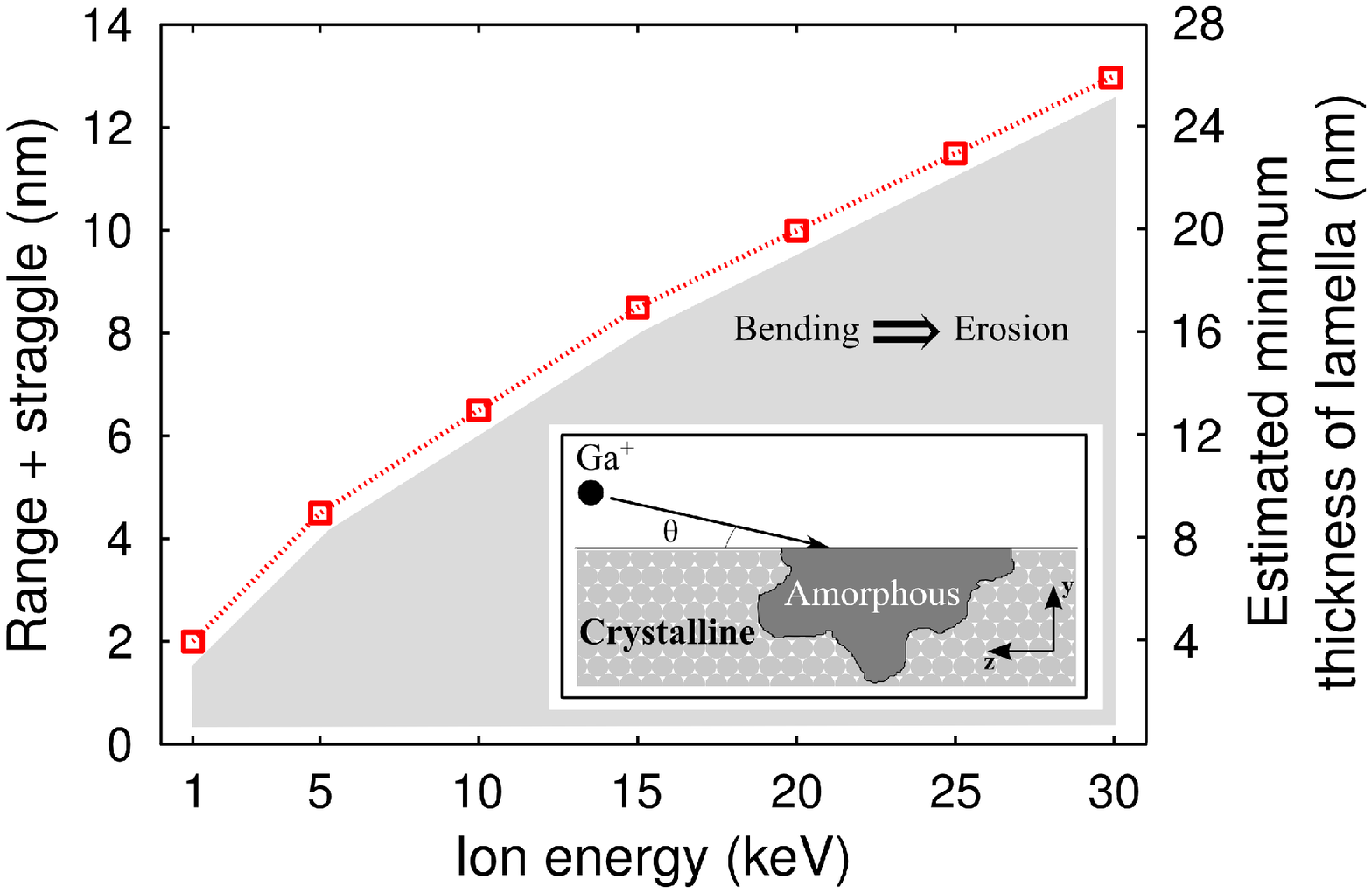}
  \caption{Sum of range and straggle in the $y$-direction from SRIM
    calculations of Ga ions incident on Si at an angle of $1 ^\circ$
    as a function of ion energy. On the basis of the MD simulations,
    this is estimated to give approximately half the maximum thickness
    of the amorphous layer. The inset shows the SRIM setup
    schematically with the coordinate axes as in the MD runs. Here
    $\theta = 1 ^\circ$. \label{fig:SRIM}}
\end{figure}

To predict the minimum attainable thickness of the lamella top edge
according to this erosion mechanism, we performed a set of SRIM runs
with Ga$^+$ incident on a Si target at an angle of $1 ^\circ$ at
energies from 1 to 30~keV. To quantify the effect of the ions in the
$y$-direction, we calculated the sum of ion range and straggle for the
different energies (Fig. \ref{fig:SRIM}). To first assess how these
results compare with MD data, we noted that for the 10~nm thick edge,
the thickness of the amorphous layer in our MD simulations in the
$y$-direction is 3 to 5~nm for 1~keV ions, and for 8~keV ions the
thickness already surpasses 10~nm. A comparison of these two points to
the SRIM results in Fig. \ref{fig:SRIM} shows that the SRIM results
give roughly half the thickness of the amorphous layer. The difference
is due to the fact that in the SRIM calculations only ions incident on
the front face of the structure are included, whereas in the MD runs,
ions incident on the very top of the structure are simulated as
well. Another factor possibly contributing to this discrepancy is a
recently predicted damage flow mechanism in amorphous
Si.\cite{Pothier2011} Taking into account this factor of two
difference leads to the prediction that at 30~keV, lamella edges of
thicknesses $\leq$26~nm will be subjected to bending into the beam and
hence catastrophic erosion. This explains why very thin lamella edges
were completely absent in our experiments using the 30~keV Ga ion
beam.

Regarding the warping seen in the experiment, the likely explanation
for the phenomenon is to be found from previous studies of Si wafer
curvature as brought on by ion implantation:\cite{Volkert91} As the
FIB is used to mill the face of the lamella, the surface region is
amorphized, which causes the region to expand and hence to bend the
lamella in the observed manner. The relevant scale for the effect is
much larger than that in the present MD simulations. Furthermore, the
mechanism relying on surface tension producing the bending toward the
beam is dominant in our study over volume changes due to
amorphization. The volume change of Si upon amorphization is predicted
to be positive (as in experiment) for EDIP and negative for
SW,\cite{Nord02} but nevertheless both potentials reproduce the
bending of the top part of the lamella edge in the same direction. In
our simulations, this is due to the small size of the structures; the
smaller a system is, the higher its surface area to volume ratio, and
hence any surface effects are pronounced in smaller systems.

%\section{CONCLUSIONS}

% We have above demonstrated two different detrimental mechanisms
% acting on a Si lamella under Ga irradiation in a FIB, which
% seemingly prohibit the creation of lamellae of less than $\sim$20~nm
% in thickness. The first mechanism, a geometric effect demonstrated
% by experiment, is the rounding of the edges of the lamella due to
% the finite sharpness of the beam. The second mechanism, demonstrated
% by atomistic simulations, is the erosion of the thin top edge of the
% lamella which explains the unexpected shrinkage of the thinnest
% ($\sim 10$~nm) part of the edge seen in our experiment for the
% double-wedge structure.

In conclusion, we have experimentally demonstrated the well-known fact
that a FIB milling energy of 30~keV cannot be used to create very thin
lamellae ($<\sim 20$~nm) within the conventional thinning
approach. Our simulations reveal that this limitation is a fundamental
one, being due to atomic-scale effects at the top edge of the lamella
which lead to the erosion of the structure under the ion
beam. Specifically, we predict that an edge of thickness $\le$26~nm
will be subject to catastrophic erosion through the lamella top edge
bending into the beam. This effect can be reduced by lowering the beam
energy, but the lower the energy, the more significant the effects of
the beam tails become (Appendix B). Together these two mechanisms of
shrinkage lead to the lower limit of $20$~nm for the thickness of the
lamella during conventional FIB thinning. Our results explain why the
alternative \emph{double-tilt} approach (Appendix E) is free of these
harmful effects, and can be used for preparing extremely thin samples
of good quality for current and future AC-HRTEM studies.

\section*{ACKNOWLEDGEMENTS}

The theoretical research for this study was supported by the Academy
of Finland Center of Excellence in Computational Molecular Science and
the Helsinki Institute of Physics. EH is grateful to Michael Moseler
and Karsten Albe for valuable comments. Generous grants of computer
time from the Center for Scientific Computing in Espoo, Finland are
thankfully acknowledged. The experimental research was funded by the
German Research Foundation (DFG) and the State of
Baden-W\"{u}rttemberg through the SALVE (Sub-\AA ngstr\"om Low-Voltage
Electron Microscopy) project.

\section*{APPENDIX A: CONVENTIONAL FIB THINNING METHOD}

\begin{figure} [ht]
\includegraphics[width=0.6\textwidth]{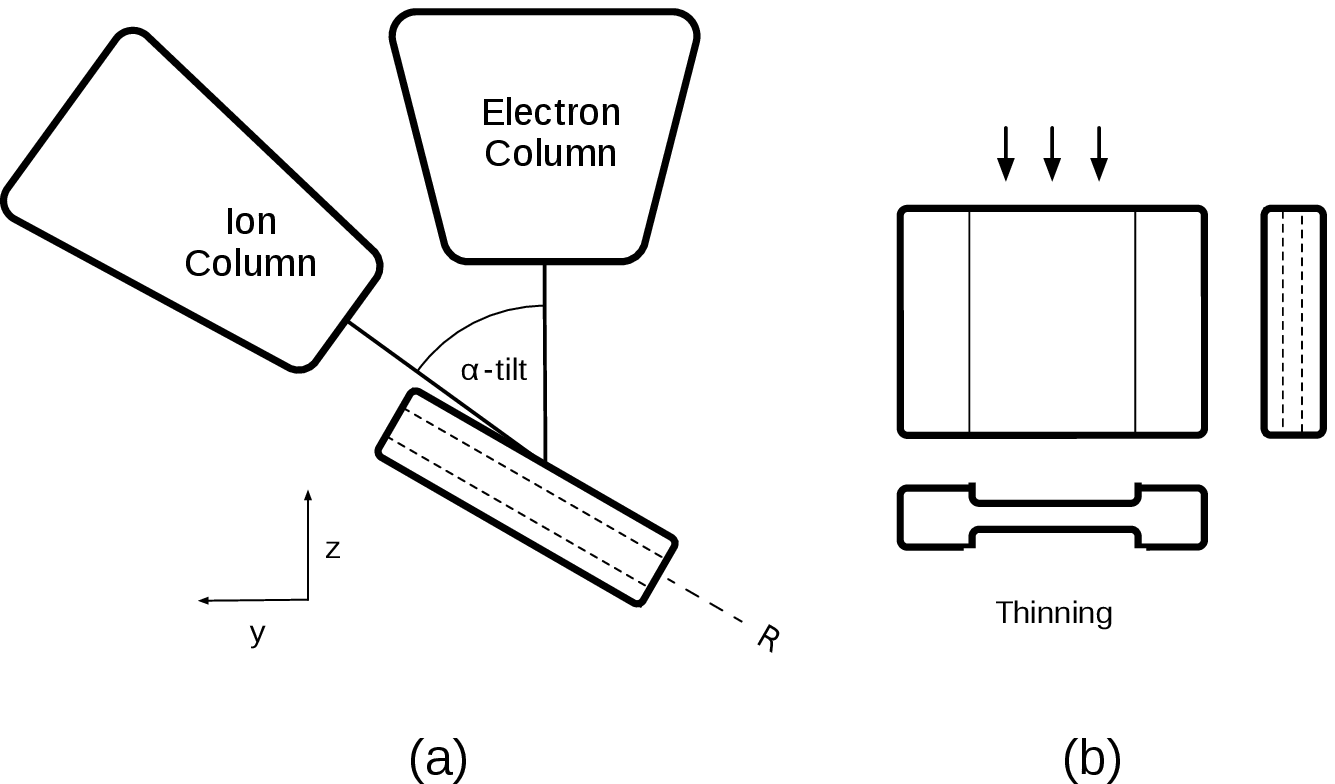}
\caption{(a) Schematic of the FIB/SEM microscope column
arrangement. (b) Orthographic third angle projection of the lamella
during conventional thinning. The direction of the Ga ions is
indicated by the arrows.
\label{fig:lamella_in_experiment}}
\end{figure}

Throughout the study, sample fabrication was performed in a Zeiss
NVision 40 CrossBeam FIB instrument incorporating a Ga liquid metal
ion source and in-situ scanning electron microscopy (SEM) imaging
using a thermal field emission source. The energy of the Ga ions can
be varied in the range from 1 to 30~keV. Figure
\ref{fig:lamella_in_experiment}(a) shows a schematic drawing of the
microscope. TEM lamella preparation by the FIB lift-out method is
carried out in four steps:\cite{Langford:2004ea} (1) Rough milling of
a lamella, (2) lift-out, (3) thinning, and (4) polishing. During rough
milling, a free-standing structure containing the region of interest
is created by removing material around it through FIB milling. Then,
for lift-out, a manipulator is attached to the structure using ion
beam induced deposition (IBID). Subsequently, the structure is cut
free from the bulk. Using the manipulator, the resulting lamella is
transferred to a special TEM lift-out grid. Once it is firmly attached
by IBID it can be thinned further using high energy ion beam
milling. In order to reduce the resulting damage layer ($\sim$30 nm
for 30 keV Ga ions in Si) the Ga ion energy is reduced for one or more
polishing step(s). Figure \ref{fig:lamella_in_experiment}(b) shows a
schematic drawing of the lamella during the thinning and polishing
process of the conventional in-situ lift-out technique. The lamella
faces are milled under a glancing angle of 1 to 3$^\circ$ (depending
on milling current) to obtain co-planar surfaces. In conventional
thinning, material is removed top--down from one or both sides of the
lamella. Polishing is done analogously, until the desired final
thickness is reached.

\section*{APPENDIX B: SPUTTERING GEOMETRY}

\begin{figure} [ht]
\includegraphics[width=0.5\textwidth]{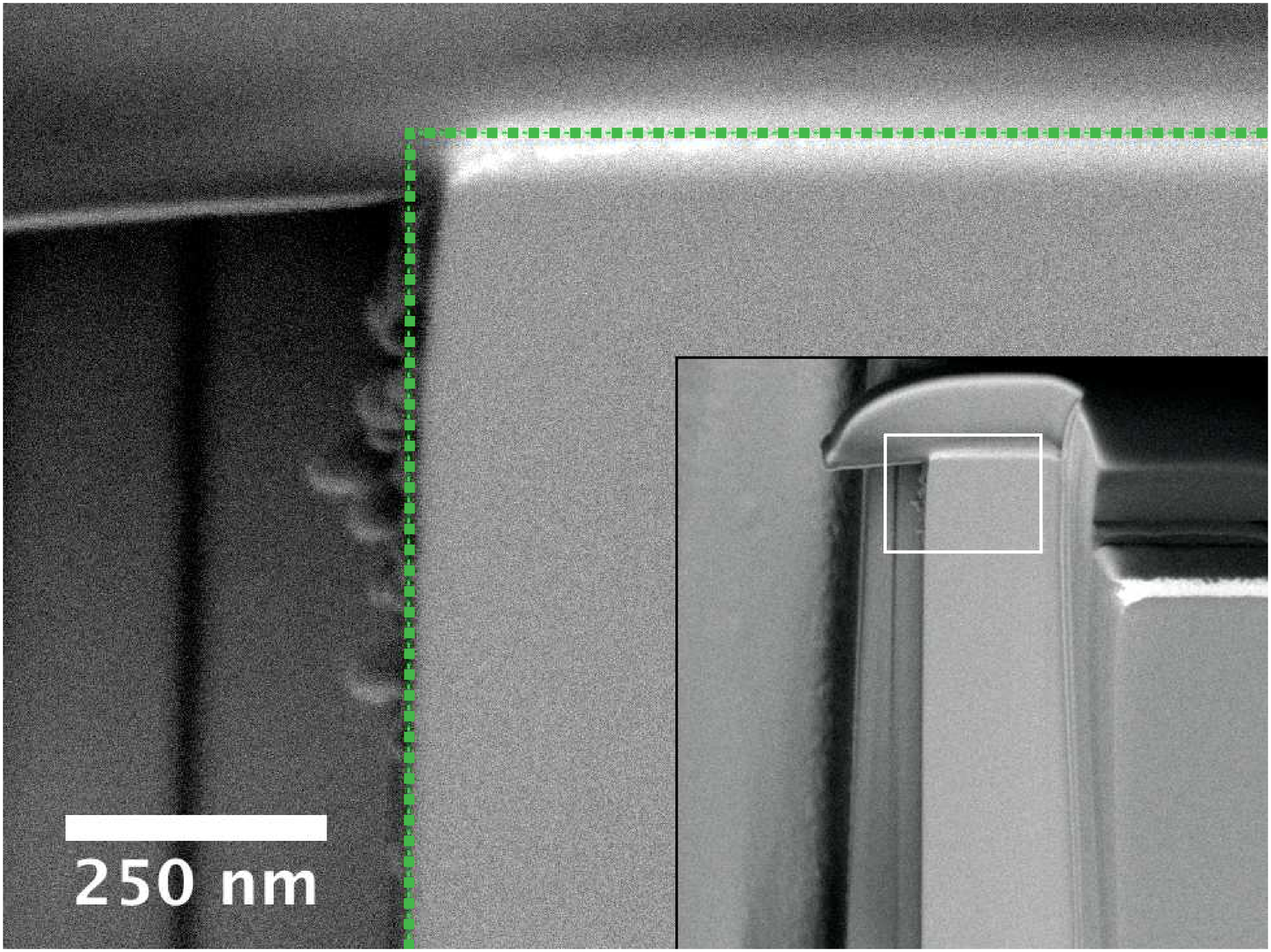}
\caption{SEM image of a cross-section through a FIB-fabricated vertical lamella sidewall.
On the top edge, rounding is
caused through milling by the ion beam tails. The (green) dotted line indicates the
ideal edge shape an infinitely sharp beam should create. Inset (black frame): Overview
of the cross-section; the (white) rectangle indicates the location of the
magnified view.
\label{fig:lamella_cross-section}}
\end{figure}

In order to understand the different effects leading to the shrinking
behavior of the lamella in the conventional FIB method, we studied the
sputtering geometry resulting from finite beam sharpness effects on a
single edge. Fig. \ref{fig:lamella_cross-section} shows an SEM image
of the cross-section through a FIB-prepared thick Si lamella. The
sidewalls of the lamella were milled using a 30~kV Ga ion beam with a
current of 10~pA at 1$^\circ$ incidence angle. The beam diameter
(FWHM) was $\sim$20 nm and the milling depth was set to 10 $\mu$m. The
beam was repeatedly scanned quickly on the same line until the desired
milling depth was reached. The resulting edge is perpendicular to the
sample surface but with a rounded top. As seen in the figure, the
rounding effect is already in the 10~nm range at a beam voltage of
30~kV, and should become much worse at lower voltages due to chromatic
lens aberrations.

\section*{APPENDIX C: SIMULATION METHOD}

As obtaining an atomic-level description of the FIB thinning process
involves following the time evolution of hundreds of thousands of
atoms for thousands of time steps for each ion impact event, MD
simulation with analytical interatomic potentials, which also provides
a reasonably accurate atomistic description, is currently the only
feasible method for the investigation. In addition to MD, we carried
out a set of calculations within the binary collision approximation
scheme as implemented in the SRIM code.\cite{SRIM-2008,SRIMbook} These
simulations were performed in order to get an estimate of how some of
our MD results could be extrapolated to larger system sizes utilizing
the simpler physical model of SRIM, as explained above.

Arguably, the most critical part in designing an MD simulation is
selecting the interaction model. For Si, more than 30 analytical
potentials have been published.\cite{Baz97} Out of these, the most
common and well-established potentials for studying radiation damage
are the Stillinger-Weber (SW),\cite{SW}, Tersoff (TS),\cite{Ter88c},
and Environment-Dependent Interatomic Potential (EDIP).\cite{Jus98}
Regarding irradiation effects, the displacement threshold energy $T_d$
and the sputtering yield $Y$ can be considered to be the decisive
properties in describing the relevant processes. Unfortunately, all of
the three mentioned potentials have been shown to underestimate $T_d$
as compared to density-functional theory
calculations.\cite{Holmstrom09} SW produces the highest average $T_d$
of 29 eV, whereas EDIP gives the lowest value of 16 eV. TS falls
between these two extremes with 19 eV. The DFT value is as high as
35--36~eV. Additionally, $Y$ given by EDIP has been shown to give the
closest match with experiment for Ar ions with energies up to
20~keV,\cite{Samela2007} whereas SW and TS agree well with each other
but somewhat underestimate the experimental results. Thus, as SW gives
the best value for $T_d$, whereas EDIP describes sputtering in best
accordance with experiments, we started our simulations using EDIP and
SW in parallel. However, it turned out that there were no significant
differences in the results between these two potentials, and hence the
work was completed using EDIP.

For modeling the Ga--Si interaction, a purely repulsive ZBL
potential~\cite{ZBL} was used, because the Coulombic interaction
between the nuclei is the overwhelmingly dominating effect in
high-energy collisions, and the finer chemistry between the two types
of atoms can be neglected. Indeed, chemical effects should affect
irradiation results only when the projectile energy is very low ($<$
100 eV).\cite{Ahlgren2011} We have previously used a similar approach
to model Ga impacts on graphene in the keV energy
range.\cite{Lehtinen2011} To realistically model high-energy Si--Si
collisions brought on by the up to 30~keV ion impacts, a separate ZBL
high-energy part was smoothly splined to the repulsive part of the
Si--Si potentials at small interatomic distances.\cite{Nor94}
Finally, to account for electronic stopping of energetic atoms, a
frictional, velocity-dependent drag force was applied to all atoms
with a kinetic energy higher than 5~eV.

To model the dissipation of energy deposited into the irradiated
structure by an incident ion, the system was coupled to a Berendsen
thermostat~\cite{Ber84} at a temperature of 10~K with a time constant
of 300~fs in a region with a thickness of 0.8~nm at the periodic
boundaries and 1.5~nm from the bottom of the structure. As those parts
of the system which are coupled to the thermostat constitute an
effective boundary in the periodic $x$-direction, the system was
translated over the periodic boundaries for each ion impact event so
that the impact point in the $x$-direction was always halfway between
the regions where temperature was being scaled. Also, to prevent the
entire structure from moving as a result of the ion impacts, a small
segment encompassing $\sim$3000 atoms in the lower back face of the
system was fixed. Each ion impact simulation was carried out for
15~ps, which was enough to model the ballistic phase of the
irradiation event. After this, the temperature of the system was
slowly decreased to the initial simulation temperature of 10~K before
the next impact. All simulations were performed using the {\scshape
  parcas} MD code.\cite{PARCAS}

\section*{APPENDIX D: SIMULATION RUNS}

The first set of runs consisted of simulating edges of thicknesses
3--10~nm.~\footnote{10~nm is the maximum thickness that could be
  modeled with our approach in a reasonable computation time.} In
order to simultaneously study both the role of thickness and that of
beam energy, we carried out these simulations for irradiation energies
of 1~keV and 30~keV up to a total beam dose of 500 ions. We also performed
the same set of runs using a beam energy of 500~eV, but found no
noteworthy differences to the 1~keV case.

Due to the differences observed for the 1~keV and 30~keV irradiations
(as described above), we performed another set of simulations
to investigate intermediate energies. These complementary runs with
beam energies of 8 and 15~keV were performed on the 10~nm edge. To
make sure we weren't missing any important effects due to too low a
dose, we at this point increased the total dose to 1000 ions. Finally,
we extended the earlier simulations for the 3 and 10~nm edges up to
this same dose. It is worth noting that even these 'high' doses are
far from the experimental ones. However, the simulations therefore
allow us to detect effects which may disappear too fast in the
experiment to be directly observed.

\section*{APPENDIX E: ATTAINING THE THINNEST POSSIBLE LAMELLA}

It has always been best practice to reduce the ion beam energy as the
lamella becomes progressively thinner. After all, the goal of TEM
sample preparation is to retain as much pristine material as possible
inside the lamella. Despite this, no ultra-thin lamellae have been
produced using the conventional method. The problem ultimately lies in
the dominating effect of chromatic lens aberrations at low
acceleration voltages which significantly increase the beam
diameter. During thinning and polishing, the resulting wide beam tails
result in the rounding of the edges of the lamella top. When a lamella
is thinned from both sides, these tails may overlap causing
shrinkage. Besides shrinkage, these tails cause the top edge to become
significantly thinner than the rest of the lamella. This thin edge is
eroded away through bending into the beam and sputtering from both
sides, as described above by our simulations. Reducing the beam energy
further does decrease these top edge effects (as seen in
Fig.~\ref{fig:SRIM}), but in turn also increases geometric shrinking
of the bulk of the lamella. The resulting trade-off seems to set a
fundamental lower limit of $\sim$20~nm to lamella thickness attainable
using the conventional thinning method.

However it was shown recently that the conventional limit can be
overcome using the \emph{double-tilt} method.\cite{Lechner2011} Within
this approach, the detrimental formation of a sharp edge during
thinning is omitted by milling perpendicular grooves in the front and
the back sides of the sample. The thin transparent area of the lamella
is formed where the grooves overlap. Geometric and top edge effects
are thus effectively suppressed since there is no free edge,
independent of the lamella's final thickness. In addition,
amorphization of the lamella surface can be reduced to a minimum since
no trade-off between beam energy, milling angle, and tolerable focus
spread has to be made. Even low-kV Ar ion polishing, typically
featuring beam diameters of 10 to 100 $\mu$m, can be performed at
grazing incidence. Last but not least, the mechanical stability of the
resulting lamella is much higher than that obtained with the
conventional method since the transparent window is surrounded on all
sides by a thick frame of material. These points explain why the
\emph{double-tilt} method can be used to obtain results superior to those of
the conventional method.

%

%\bibliographystyle{aipnum4-1}
%\bibliography{/home/eholmstr/papers/bib/general}
%
%\begin{thebibliography}{10}
%
%\end{thebibliography}

\end{document}